\documentclass{jfm}
\usepackage{graphicx}
\usepackage{epstopdf, epsfig}
\usepackage{color}
\usepackage{amsmath}
\usepackage{csquotes}
\usepackage{subfig}
\usepackage{graphicx}
\DeclareMathOperator*{\minimize}{Minimize}
\usepackage[linktocpage,colorlinks=true,linkcolor=blue,citecolor=blue,breaklinks=true]{hyperref}
\usepackage[normalem]{ulem}
\usepackage[bordercolor=white,backgroundcolor=gray!30,linecolor=black,colorinlistoftodos]{todonotes}

\usepackage[framemethod=tikz]{mdframed}

\shorttitle{Brachistochrone of a fluid-filled bottle}
\shortauthor{Gurram, Raja, Mahapatra and Panchagnula}
\title{\textbf{On the brachistochrone of a fluid-filled cylinder}}
\author{Srikanth Sarma Gurram\aff{1},
Sharan Raja\aff{1}, Pallab Sinha Mahapatra\aff{1} \and Mahesh V. Panchagnula\aff{2} \corresp{\email{mvp@iitm.ac.in}}}
\affiliation{\aff{1}Department of Mechanical Engineering, Indian Institute of Technology Madras, Chennai 600036 India,
\aff{2}Department of Applied Mechanics, Indian Institute of Technology Madras, Chennai 600036 India}
\begin{document}
\maketitle
\begin{abstract}
We discuss a fluid dynamic variant of the classical Bernoulli's brachistochrone problem. The classical brachistochrone for a non-dissipative particle is governed by maximization of the particle's kinetic energy resulting in a cycloid. We consider a variant where the particle is replaced by a bottle filled with a viscous fluid and attempt to identify the shape of a curve connecting two points along which the bottle would move in the shortest time. We derive the system of integro-differential equations governing system dynamics for a given shape of the curve. Using these equations, we pose the brachistochrone problem invoking optimal control formalism and show that (in general) the curve deviates from a cycloid. This is due to the fact that increasing the rate of change of bottle kinetic energy is accompanied by increased viscous dissipation. We show that the bottle motion is governed by a balance between the desire to minimize travel time and the need to reach the end point in the face of increased dissipation. The trade-off between these two physical forces plays a vital role in determining the brachistochrone of a fluid-filled cylinder. We show that in the two limits of either vanishing or high viscosity, the brachistochrone for this problem reduces to a cycloid. An intermediate viscosity range is identified where the fluid brachistochrone is non-cycloidal. Finally, we show relevance of these results to the dynamics of a rolling liquid marble.
\end{abstract}
\section{Introduction}
In 1696, Johann Bernoulli posed a famous problem to the readers of Acta Eruditorum: ``{\em{Given two points $A$ and $B$ in a vertical plane, what is the curve traced out by a point (particle) acted on only by gravity, which starts at $A$ and reaches $B$ in the shortest time?}}" \citep{andre1696acta}. The solution of this problem was shown to be a cycloid connecting $A$ and $B$ and was probably responsible for the inception of calculus of variations, which has had a significant impact on the evolution of modern science and engineering. We are interested in a fluid dynamic variant of this question where the point particle is replaced by a fluid-filled cylinder. The problem is interesting in that the system dynamics is a function of the time history of angular velocity of the cylinder. This memory dependence of the system is what is likely to make the brachistochrone of a fluid-filled cylinder different from a cycloid. This property has been illustrated by \citet{supekar2014dynamics} in the context of discussing the dynamics of a fluid-filled cylinder rolling down a straight inclined plane.

The classical brachistochrone problem has been extended by previous researchers. For example, \citet{extbrac} introduced field varying gravity, \citet{optcontrol} included the effect of coulomb friction between the curve and the particle and \cite{cherkasov2017range} includes both coulomb and viscous friction. Most relevantly, \cite{diss} included the effect of a non-conservative fluid drag force in a simple Stokesian form (as though the point particle was moving in a universe of a viscous fluid). \cite{supekar2014dynamics} discussed the full effects of fluid dynamics on the motion of a fluid-filled cylinder moving down an inclined plane.

Despite the long history of the brachistochrone problem, it has not been extended formally into the realm of fluid dynamics except by modeling drag effects phenomenologically. We attempt to close this gap in the literature by invoking the coupled equations of fluid dynamics and rigid body mechanics to calculate the brachistochrone for a fluid-filled cylindrical shell (herein referred to as the bottle) with the fluid viscosity being non-negligible.  
%
Specifically, we ask the following question.
\begin{displayquote}
\textit{Given two points A and B in a vertical plane, what is the curve traced out by the center of mass of a fluid-filled bottle acted on only by gravity, on which the center of mass starts at A and reaches B in the shortest time?}
\end{displayquote}

The answer to this question will have a broader impact on a wide range of fluid dynamic problems. Problems where a performance measure (say, total flow rate) of a fluid dynamical process is to be optimized while accounting for viscous dissipation are potential candidates which are likely to benefit from the approach discussed herein. For example, consider a related problem where one desires to calculate the shape of an axisymmetric smooth funnel of known inlet and outlet cross-sectional circular areas and of a fixed height difference between the inlet and outlet cross-sections, such that the funnel has a maximum rate of gravitational drainage of a fluid. It is well known that flow through a funnel (or, analogously, through a pressure swirl atomizer) exhibits multiple states as has been discussed by \citet{Taylor1948,Taylor1950} and  \cite{Binnie1950}. This is due to the three-dimensional nature of the boundary layer \citep{Bloor1977}. Therefore, the brachistochrone funnel shape is likely to be a non-trivial function of the fluid dynamic parameters. A second related fluid brachistochrone problem could involve calculating the shape of a `water slide' connecting two points which maximizes the flow rate. We are motivated in this work to lay the general optimal control based mathematical framework to obtain solutions to such problems.

\section{Problem formulation}
\begin{figure}
\begin{center}
\includegraphics[width=\textwidth,trim={0cm 0cm 0cm 0cm},clip]{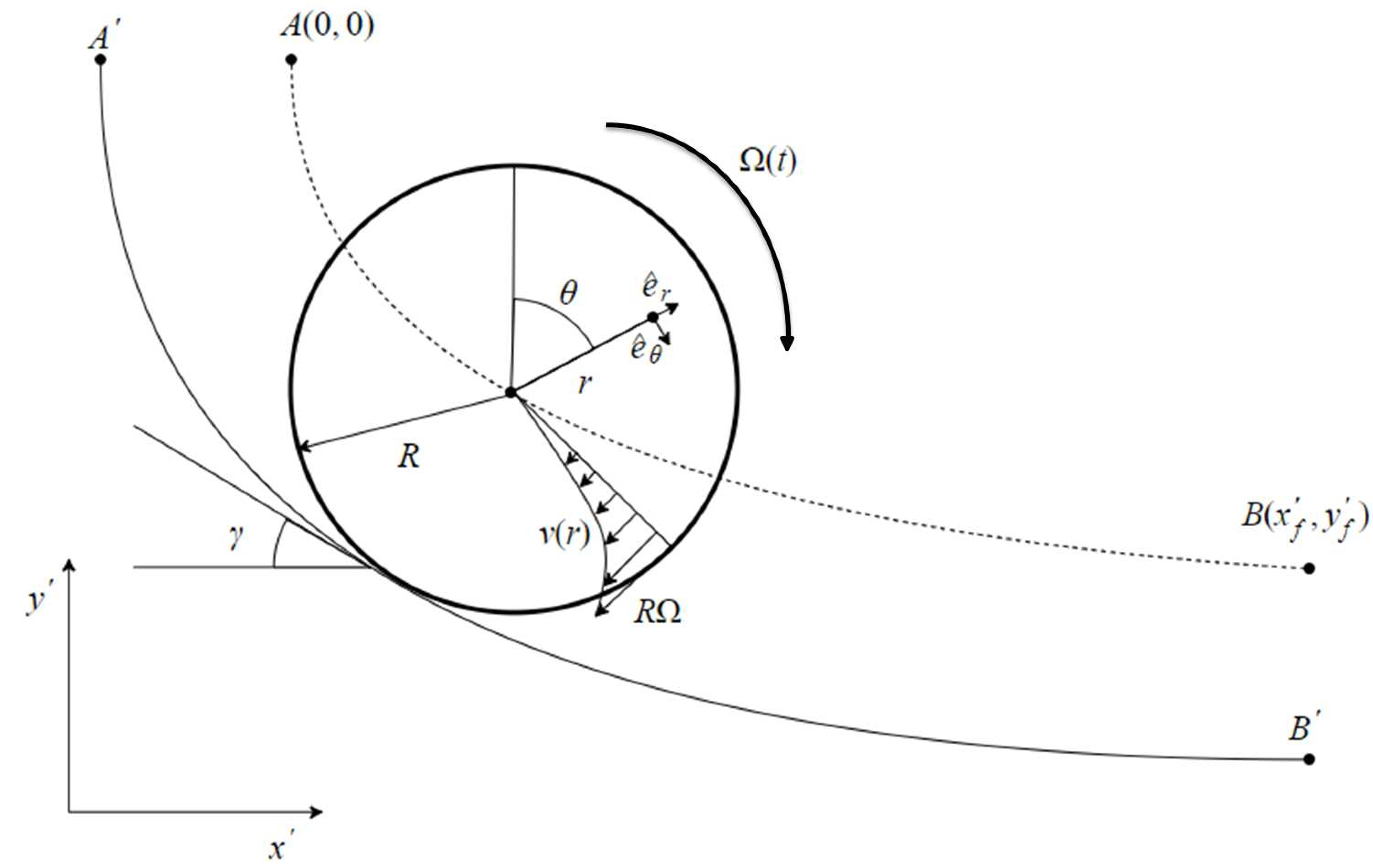}
\end{center}
\caption{Schematic of the problem statement. The fluid-filled bottle is released from $A$ and allowed to roll towards $B$ subject to acceleration due to gravity in the $-y'$ direction. $\hat{e}_r$ and $\hat{e}_\theta$ define a local co-ordinate system fixed to the center of mass and translating with it. An illustrative instantaneous nonlinear azimuthal velocity profile is indicated for the fluid inside the bottle. The curve shown in solid line is the surface on which the cylinder would roll without slipping while the curve shown in dotted line is the locus of the center of mass. The surface (solid line) is offset by a factor of $R$ from the locus of the center of mass (dotted line).}
\label{fig:schematic}
\end{figure}

Consider a bottle comprising of a thin cylindrical shell of radius $R$, length $h$ and mass $M$ fully filled with a fluid of kinematic viscosity $\nu$ and density $\rho$. Now, consider such a bottle rolling down an inclined curved path $y'=f(x')$ from a point $(0,0)$ (without loss of generality) to an end point $(x'_f,y'_f)$ as shown in figure \ref{fig:schematic}. The dynamics of this body are governed by the Navier-Stokes equations for the fluid and the equations of rigid body motion for the shell. Appropriate interface conditions couple these two sets of equations. In the following section, we derive these governing equations.
\subsection{Governing equations of dynamics}
The axisymmetric motion of a fluid due to a rotating cylinder was studied by \cite{batchelor} in the usual polar coordinate $(r,\theta)$ description. The corresponding equations written in the stationary frame of reference fixed at the center of the cylinder are
\begin{subequations}
\begin{equation}
\frac{\rho v^2}{r} = \frac{\partial p}{\partial r},
\label{ge1}
\end{equation}
\begin{equation}
\frac{\partial v}{\partial t} = \nu \frac{\partial}{\partial r}
\left(\frac{1}{r} \frac{\partial}{\partial r}\left(rv\right)\right).
\label{ge2}
\end{equation}
\label{ge}
\end{subequations}

\noindent In the above equations, $v$ is the azimuthal velocity of the fluid and $p$ is the pressure field. If the center of mass $G$ is accelerating with an acceleration $\mathbf{a}$, a superposed pressure field $p_s$ results due to the d'Alembert force. This field is given by $\nabla p_s=\rho \mathbf{a}$. However, the velocity field would remain unaltered.


The corresponding equations of motion of the bottle are
\begin{subequations}
\begin{equation}
(m+M)R\dot{\Omega} = (m+M)g \sin{\gamma(t)}-f,
\label{fbd1}
\end{equation}
\begin{equation}
MR^2\dot{\Omega} = fR-T.
\label{fbd2}
\end{equation}
\label{rbe}
\end{subequations}

\noindent Here, $m=\rho \pi R^2h$ is the mass of the fluid contained in the shell and $f$ is the contact friction. In addition, $T$ is the torque exerted on the shell wall by the fluid as a result of the wall shear stress and is given by 
\begin{equation}
T= 2\pi \mu  R^2h \left.\left( \frac{\partial v}{\partial r}-\frac{v}{r} \right) \right|_{r=R}.
\label{ShearStress}
\end{equation}
\noindent It is to be noted that $p_s$ does not contribute to the torque exerted by the fluid on the wall. The boundary conditions for equations \eqref{ge} are
\begin{subequations}
\begin{equation}
v(R,t)=R \Omega(t),
\label{cBC}
\end{equation}
\begin{equation}
\frac{\partial v}{\partial r}(0,t)=0.
\end{equation}
\end{subequations}

\noindent It can be seen that equation \eqref{cBC} provides the necessary coupling between the fluid motion and that of the shell. Since the fluid bottle is assumed to be starting from rest and the fluid is assumed to be quiescent initially,
\begin{equation}
\Omega(0)=0 \qquad \textrm{and} \qquad v(0,r)=0.
\label{ic}
\end{equation}
Equations \eqref{ge} to \eqref{ic} pose an initial boundary value problem (IBVP) which has been solved closed form by \citet{batchelor} for a constant angular velocity condition $(\Omega(t)=\Omega_0)$. In contrast, the angular velocity of the cylinder in our case evolves in time, which is governed by equations \eqref{rbe}. \citet{supekar2014dynamics} have extended the closed form solution of \citet{batchelor} to the case of a time varying angular velocity using Duhamel's theorem. They studied the case of a cylinder rolling down an inclined plane and showed that the velocity field is given by

\begin{equation}
v(r,t) = 
\int_{0}^{t}\left( r+2R\sum_{n=1}^{\infty}\frac{J_1\left(\lambda_n\frac{r}{R}\right)}{\lambda_n J_0(\lambda_n)}
\exp{\left(-\lambda_n^2 \frac{\nu (t-\tau)}{R^2}\right)}
\right) \dot{\Omega}(\tau) d\tau.
\label{VelocityEquation}
\end{equation}

\noindent A brief discussion of the formal extension of Batchelor's solution to the present case can be found in Appendix \ref{labelB}. We extend their solution to the case where the fluid-filled bottle encounters varying angles of inclination along the curved path. For our case, using equation \eqref{VelocityEquation}, equations \eqref{rbe} reduce to

\begin{subequations}
\begin{equation}
(m+2M) R^2\dot{\Omega}(t) = -T(t) + (m+M) g R sin(\gamma(t)),
\label{t1}
\end{equation}
\begin{equation}
T(t) = 4\pi \nu \rho R^2 h \sum_{n=1}^{\infty}{\int_{0}^{t}{exp{\left(-\lambda^2_n\frac{\nu (t-\tau)}{R^2}\right)}\dot{\Omega}(\tau)}}d\tau.
\label{t2}
\end{equation}
\label{TorqueEquation}
\end{subequations}

\noindent Here, $\lambda_n$ is the $n^{th}$ root of the Bessel function of first kind. We choose to non-dimensionalize the above equations with $\frac{R^2}{\nu}$ as the time scale, $R$ as the length scale and $\rho \pi R^2 h$ as the mass scale. The resulting non-dimensional equations are:
\begin{subequations}
\begin{equation} \label{eq 1}
(1+2\pi_m)\dot{\Omega}(t) = -4 \mathcal{T}(t) + (1+\pi_m)\pi_g \sin(\gamma(t)),
\end{equation}
\noindent where,
\begin{equation} \label{eq 2}
\mathcal{T}(t) = \sum_{n=1}^{\infty}{\int_{0}^{t}{exp(-\lambda^2_n(t-\tau))\dot{\Omega}(\tau)}}d\tau.
\end{equation}
\label{ndeqns}
\end{subequations}

\noindent In addition, the equations governing the kinematics of the center of mass of the fluid bottle are
\begin{subequations}
\begin{equation}
\dot{x}(t) = \Omega(t) \cos\gamma(t),
\end{equation}
\begin{equation}
\dot{y}(t) = -\Omega(t) \sin\gamma(t).
\end{equation}
\label{eq 3}
\end{subequations}

\noindent Two non-dimensional parameters are identified to govern the dynamics of the fluid-filled bottle. They are
\begin{align}
\pi_m = \frac{M}{m} \qquad \textrm{and} \qquad \pi_g=\frac{g R^3}{\nu^2}.
\label{nd}
\end{align}
Here,  $\pi_m$ is the ratio of the cylinder mass to the fluid mass and is a measure of the inertia of the two masses and $\pi_g$ is a measure of the ratio of the viscous time scale to the gravitational time scale. In addition, the co-ordinate variables $(x,y)$ are taken to be dimensionless as well written in units of $R$. Finally, we define $(x_f,y_f)=(\frac{x'_f}{R},\frac{y'_f}{R})$ as the dimensionless terminal point of the brachistochrone. It may be noted that we are interested in calculating the trajectory traced by the center of mass of the fluid-bottle system. The actual inclined curve over which the bottle moves is obtained by offsetting the calculated curve by a distance $R$.

\subsection{Variational formulation of the brachistochrone problem}

We now formulate the brachistochrone problem in an optimal control framework as discussed in \cite{Liberzon}. In this framework, the position $\big(x(t),y(t)\big)$ and angular velocity $\Omega(t)$ of the fluid bottle comprise the state space and the instantaneous inclination of the candidate curve $\gamma(t)$ is the control action which controls the trajectory of the fluid bottle. The problem is posed to determine the control action $\gamma(t)$ which will minimize the total time of descent ($t_f$) of the fluid bottle from a point A to point B. Mathematically, the corresponding objective function can be written as

\begin{equation}
\minimize_{\gamma(t) \in \left[-\frac{\pi}{2}\textrm{, }\frac{\pi}{2}\right]} J(\gamma(t))=\int_0^{t_f} dt,
\label{opt}
\end{equation}

\noindent such that

\begin{equation}
\begin{array}{l}
x(0)=x_0,\qquad
y(0)=y_0,\\
x(t_f)=x_f,\qquad
y(t_f)=y_f.
\end{array}
\label{const}
\end{equation}

\noindent Equation \eqref{opt} along with equations \eqref{ndeqns}, \eqref{eq 3} and \eqref{const} as constraints results in a well-posed optimal control problem for determining the brachistochrone for a fluid-filled bottle. We use direct methods of optimal control to numerically solve this optimization problem. In this approach, we discretize the equations \eqref{ndeqns} to \eqref{const} using an explicit forward difference scheme in time. It must be recalled that the velocity field is known closed form and its effect on the motion of the fluid bottle is completely described by equation \eqref{eq 2}.

\section{Results}
\begin{figure}
\begin{center}
\centering
\captionsetup[subfigure]{labelformat=empty,labelsep=none}
\subfloat[(a)]{\includegraphics[width=\textwidth]{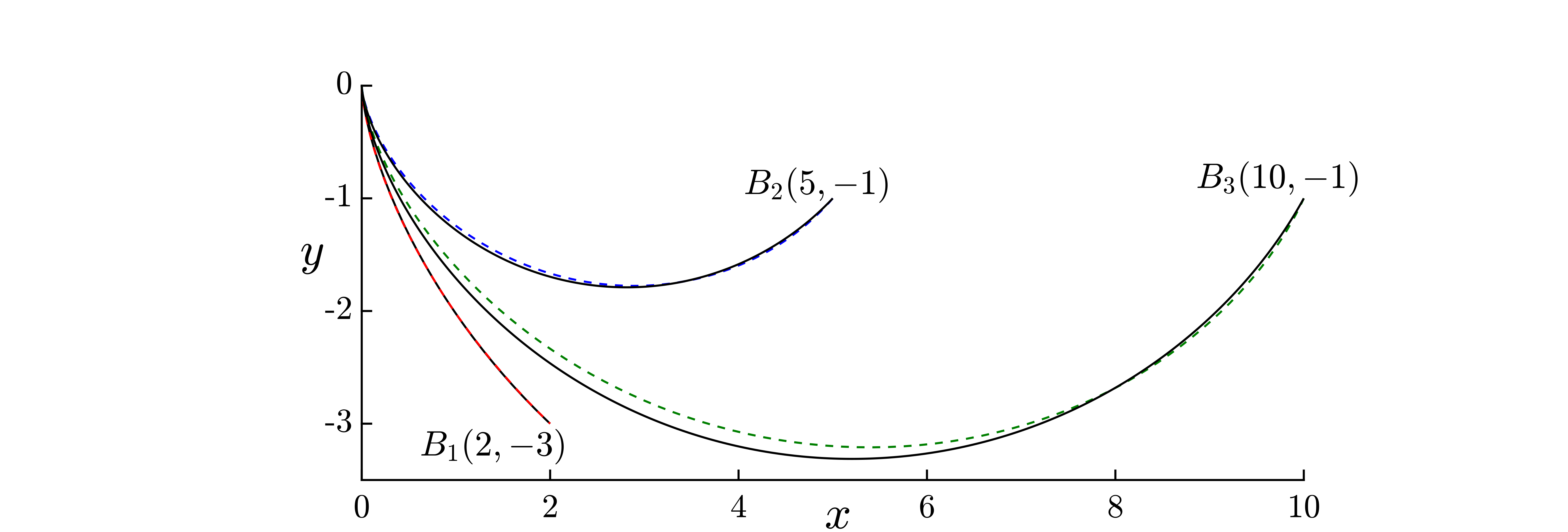}}\\
\subfloat[(b)]{\includegraphics[width=0.8\textwidth,clip]{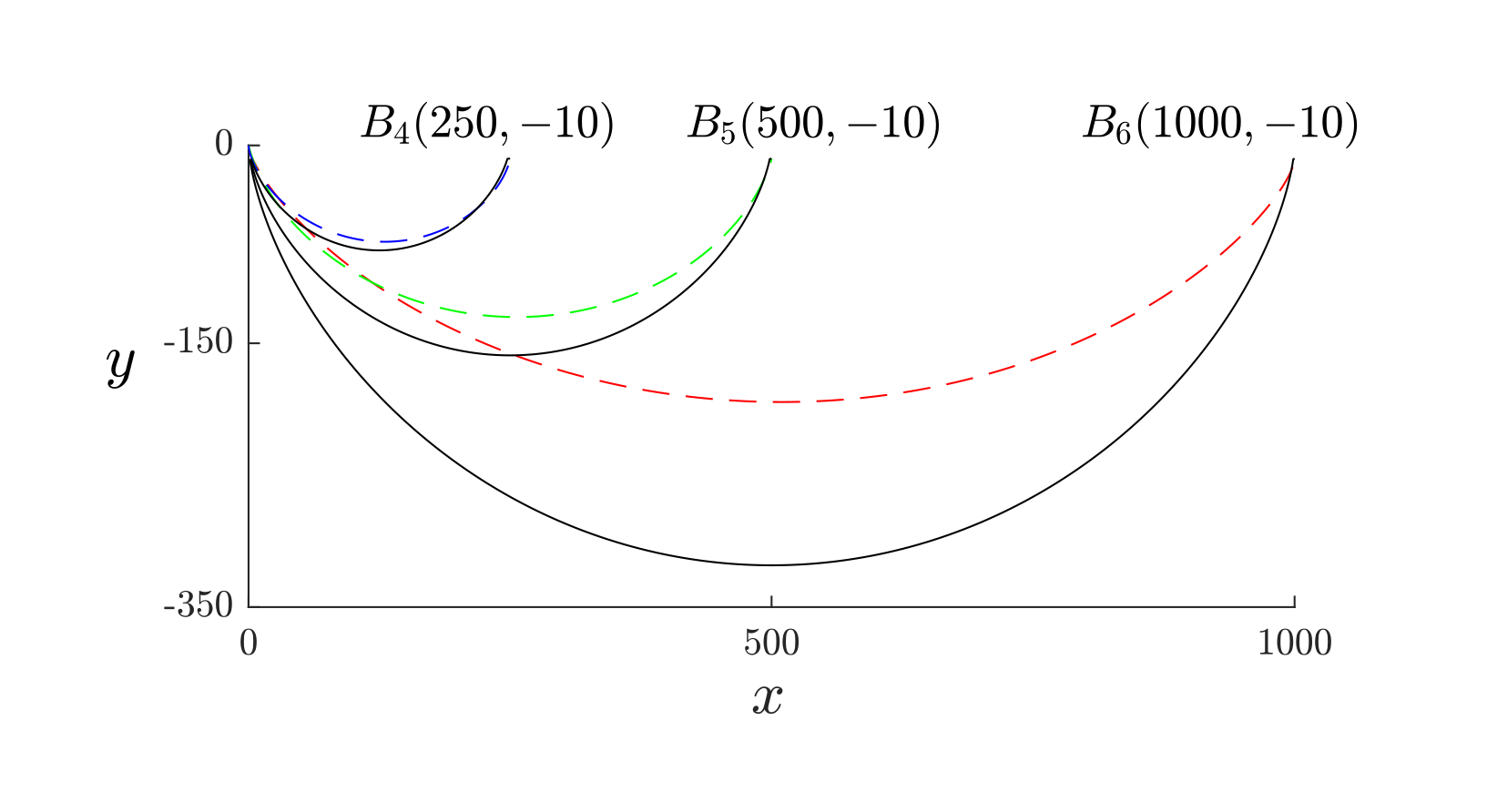}}
\caption{Figure (a) shows brachistochrone curves (shown in dotted lines) computed for terminal points $(x_f,y_f)$ at $B_1(2,-3)$, $B_2(5,-1)$ and $B_3(10,-1)$ and $\pi_g=10^3$ and $\pi_m=0.1$. The solid black curve in each case is the brachistochrone for a non-dissipative point mass i.e., a cycloid passing through the origin and the terminal point. Figure (b) is similar to (a) except for $(x_f,y_f)$ at $B_4(250,-10)$, $B_5(500,-10)$ and $B_6(1000,-10)$ The brachistochrone of the fluid bottle shows a significant deviation from a cycloid for large $x_f$ (see Figure (b)).}
\label{fig2}
\end{center} 
\end{figure}

\begin{figure}
\begin{center}
\centering
\captionsetup[subfigure]{labelformat=empty,labelsep=none}

\subfloat[(a)]{\includegraphics[width=0.5\textwidth,trim={13cm 0cm 13cm 0cm},clip]{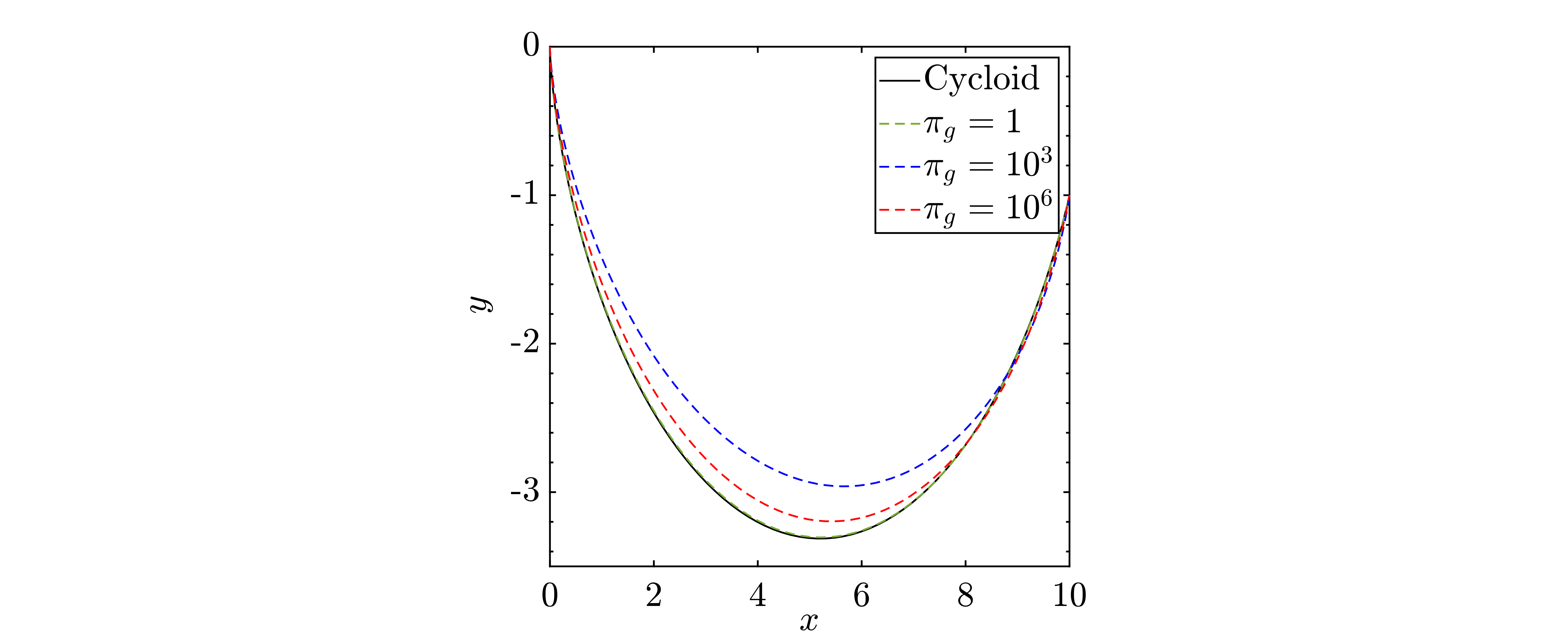}}
\subfloat[(b)]{\includegraphics[width=0.5\textwidth,trim={12.2cm 0cm 12.2cm 0cm},clip]{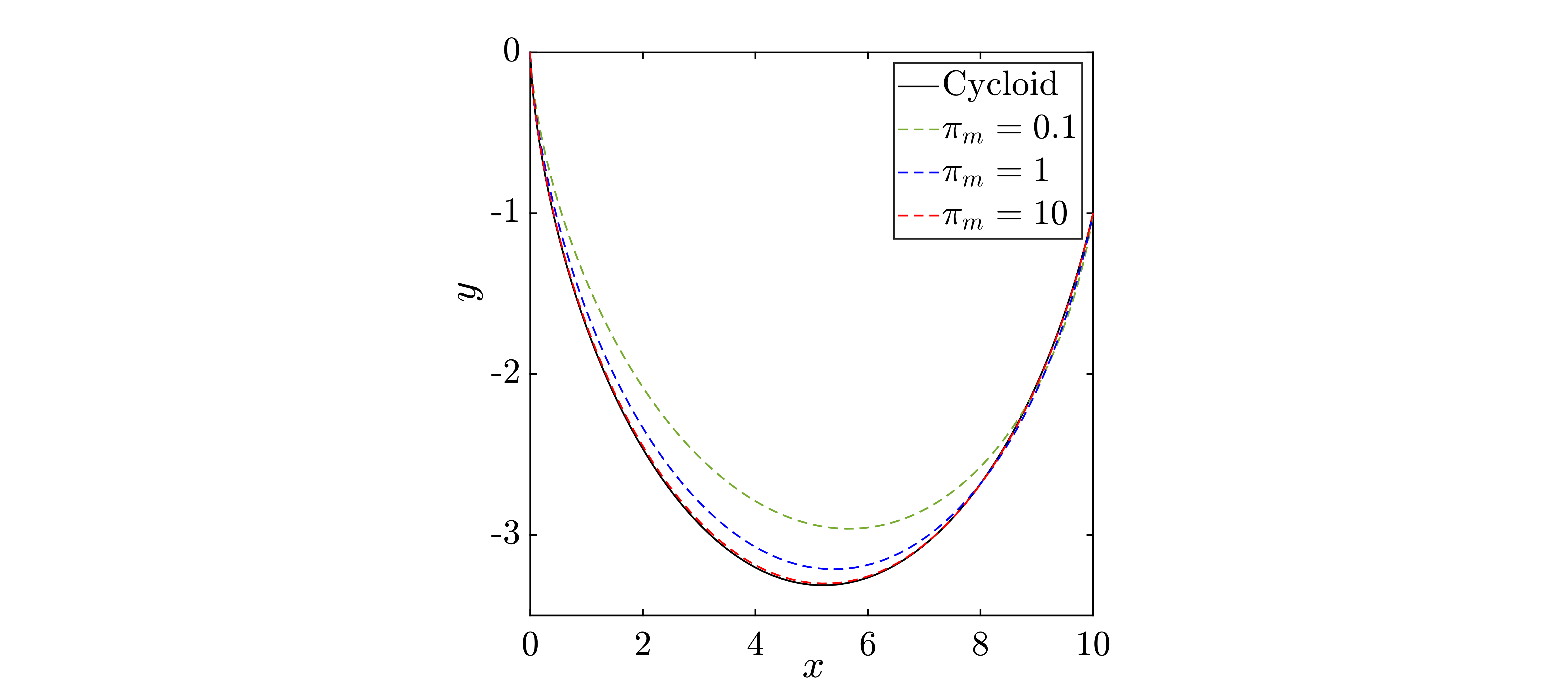}}

\caption{Plots (a) and (b) show the variation of the brachistochrone with $\pi_g = \frac{g R^3}{\nu^2}$ at $\pi_m = 0.1$ and $\pi_m = \frac{M}{m}$ at $\pi_g = 10^3$ respectively for $(x_f,y_f)=(10,-1)$. The curves vary monotonically with $\pi_m$ whereas the behaviour is non-monotonic with respect to $\pi_g$.}
\label{fig3}
\end{center} 
\end{figure}

We will begin with a discussion of the overall system dynamics and the competing objectives governing the optimization problem. In the classical Bernoulli's problem where mechanical energy is conserved, the only objective is to minimize time taken to travel from the start point (say $A$) to an end point (say $B$). In this case, the brachistochrone is governed by an interplay between maximization of kinetic energy and minimization of distance traversed. In contrast, the fluid dynamic variant considered herein is a non-conservative system. Therefore, a third competing objective arises where the fluid-filled bottle needs to conserve energy to reach the end point, while still maximizing kinetic energy. The nature of this objective is best understood by considering a case of large $x_f$. Under this condition, the fluid-filled bottle is required to (first) reach the end point in spite of viscous dissipation. We will first discuss the geometry of the brachistochrone in relation to a cycloid as a function of $x_f$, for both small as well as large $x_f$.

Figure \ref{fig2} shows the brachistochrone for these two sets of values of $x_f$. Figure \ref{fig2}(a) shows the brachistochrone curves for three different $x_f$ in the small $x_f$ regime. All brachistochrone curves begin at $(0,0)$ and are shown as dotted lines. The solid lines are plots of a cycloid passing between the same two end-points. As can be seen in this figure, the brachistochrone is different from the cycloid for end-points $B_2(5,-1)$ and $B_3(10,-1)$ while it is almost coincident for end point $B_1(2,-3)$. In addition, one can observe that the initial slope of curve near $(0,0)$ is steeper for the cycloid than for the brachistochrone of the fluid-filled bottle. This is a result of restraining effect of the drag torque in the fluid-filled bottle.
For end-points $B_2$ and $B_3$, the brachistochrone requires that the bottle as well as the fluid inside accelerate and then decelerate. In contrast, for end-point $B_1$, the fluid as well as the bottle are only required to accelerate. As a consequence of this physics, fluid dynamic memory effects characterized by the integral in equation \eqref{eq 2} begin to play a dominant role. Therefore, for a given starting point, as $y_f$ decreases while maintaining $x_f$ constant, the brachistochrone tends to approach a cycloid. 

Figure \ref{fig2}(b) shows the brachistochrone curves for three different $x_f (= 250, 500, 1000)$ in the large $x_f$ regime. $y_f=-10$ in all three cases. As can be seen, the brachistochrone curves show increasing deviation from a cycloid as $x_f$ increases. The undershoot in the brachistochrone is also lower because the objective of the cylinder (for large $x_f$) is increasingly focused on just reaching the end point (while still minimizing the time of travel). 

\subsection{Effect of $\pi_m$ and $\pi_g$}

We will now discuss the effect of the two dimensionless parameters, $\pi_m$ and $\pi_g$ defined in equation \eqref{nd}, on the geometry of the brachistochrone. Figure \ref{fig3}(a) shows the brachistochrone curves between $A(0,0)$ and $B(10,-1)$ for varying values of $\pi_g$ while $\pi_m$ was maintained constant at $0.1$.  As $\pi_g$ increases, the brachistochrone begins to deviate from a cycloid. Beyond a critical value, it begins to approach the cycloid again. In other words, the effect of $\pi_g$ is non-monotonic. This is because of the physics at the asymptotic limits of $\pi_g \to 0, \infty$. $\pi_g \to 0$ is equivalent to $\nu \to \infty$. Under this condition, the system behaves like a rigid cylinder, where it is well known that the brachistochrone is a cycloid. When $\nu \to 0$, again, the fluid plays no role and the brachistochrone coincides with a cycloid. While mathematically allowed, one must observe caution in taking the $\nu \to 0$ limit, since non-axisymmetric instabilities could play a role in determining the flow field. A mathematical analysis of these asymptotic limiting cases is included in Appendix \ref{labelA}.

 Figure \ref{fig3}(b) shows the brachistochrone curves between the same two points for varying values of $\pi_m$ while $\pi_g$ is maintained constant at $10^3$. $\pi_m>1$ indicates that the fluid's mass is greater than that of the bottle.  Not surprisingly, as $\pi_m \to 0$, the brachistochrone coincides with a cycloid because the bottle inertia is going to dominate system dynamics. As $\pi_m$ increases, the brachistochrone curves show increasing deviation from a cycloid.

It would be interesting to identify the range of $\pi_m$ and $\pi_g$ values where the brachistochrone for a fluid bottle is likely to deviate from a cycloid passing through the same start and end points. In order to quantify this deviation, a geometric norm $\delta$ is defined as
\begin{equation}\label{deviation}
\delta = \int_{0}^{x_f} \left|f(x) - f_{c}(x)\right| dx.
\end{equation}
\noindent Here, $\delta$ is the area contained between the actual brachistochrone for the fluid bottle, $y=f(x)$, and a cycloid passing through $A(0,0)$ and $B(x_f,y_f)$, $y=f_{c}(x)$. In order to quantify a performance deviation, we also define a normalized time $T^*$ as a second measure. $T^*$ is a ratio given by 
\begin{equation}
T^*=\frac{t_f}{t_c},
\label{kdev}
\end{equation}
\noindent where, $t_f$ is the total travel time of the fluid bottle on the brachistochrone between two points $A$ and $B$. $t_c$ is the time taken by a point particle on a cycloid passing through the same two points. This definition of $T^*$ is motivated by a desire to quantify the fluid's role in determining the overall objective of a brachistochrone, which is to transport a bottle between the two points in the fastest time. Since the fluid always exerts an opposing drag (see equation \eqref{eq 1}), $T^*$ is always greater than $1$. As discussed before, $\delta \to 0$ and $T^*$ asymptotes to constant values when $\pi_g \to 0, \infty$ (both for high and low viscosity fluids). From the above discussion, it is to be anticipated that at particular values of $\pi_g$, $\delta$ and $T^*$ would exhibit maxima. We are interested in the behavior of the system near this value of $\pi_g$ where $\delta$ is maximum, since it is near that parametric value where fluid effects are most significant. 

\begin{figure}
\begin{center}
\centering
\captionsetup[subfigure]{labelformat=empty,labelsep=none}

\subfloat[(a)]{\includegraphics[width=0.5\textwidth,trim={12.2cm 0cm 12.2cm 0cm},clip]{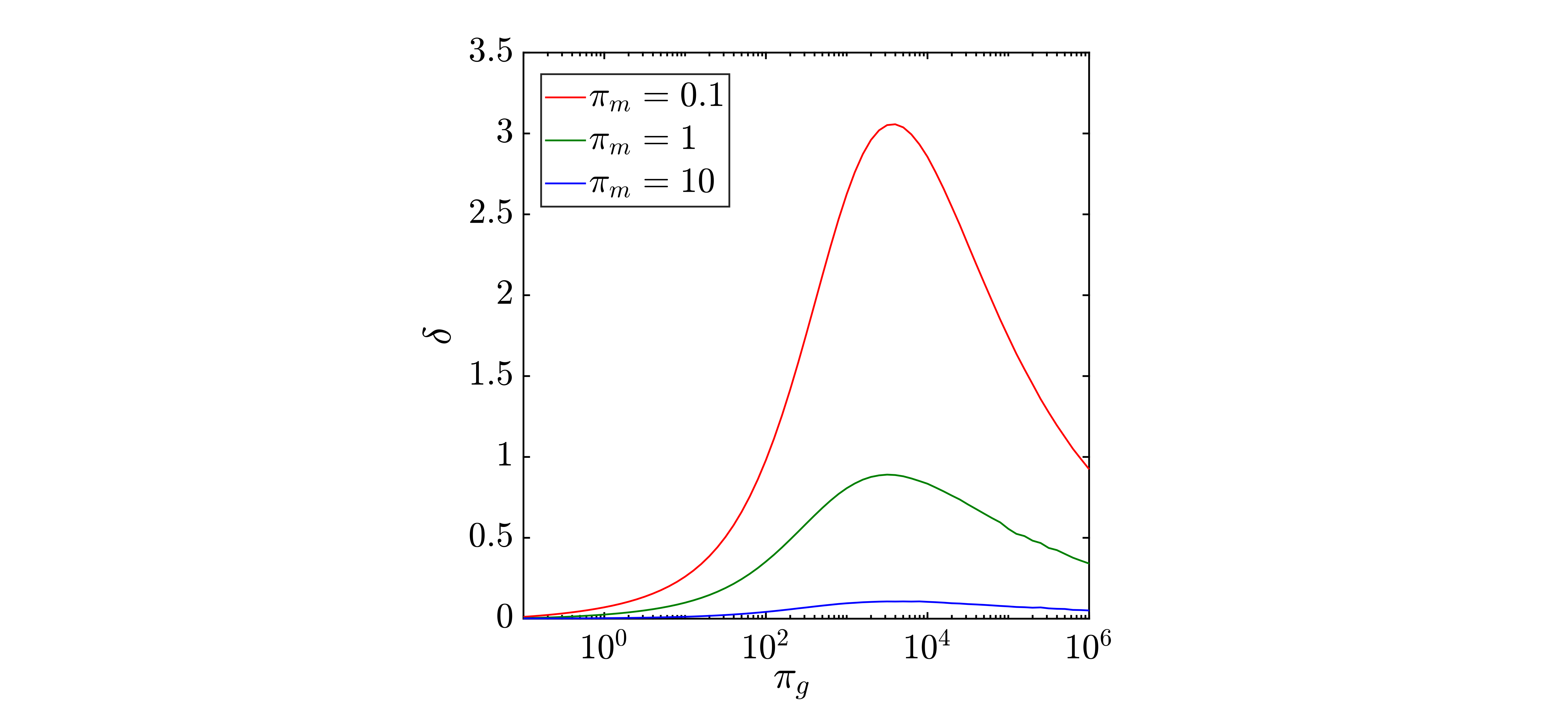}}
\subfloat[(b)]{\includegraphics[width=0.5\textwidth,trim={9cm 0cm 9cm 0cm},clip]{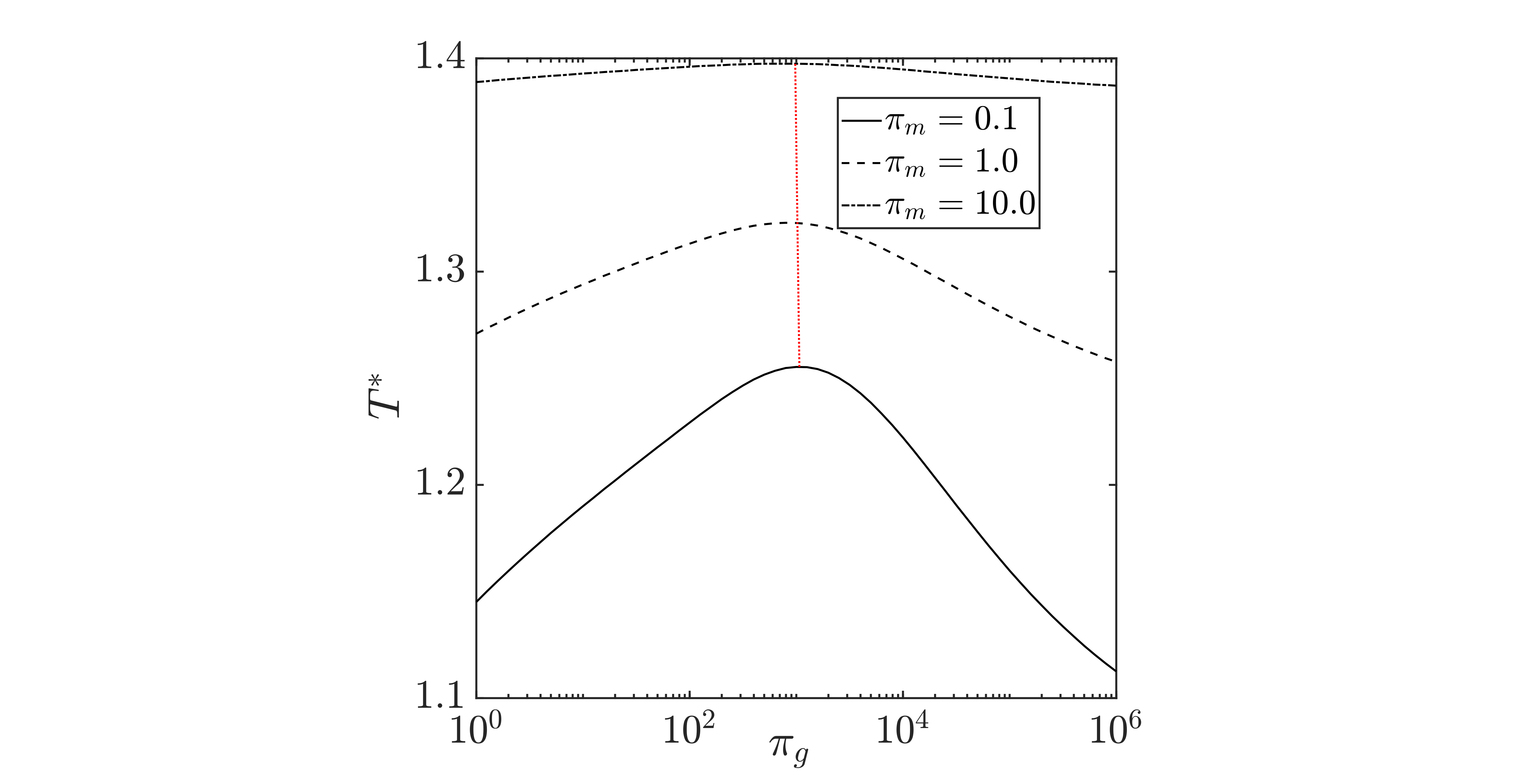}}\\
\subfloat[(c)]{\includegraphics[width=0.5\textwidth,trim={13cm 0cm 13cm 0cm},clip]{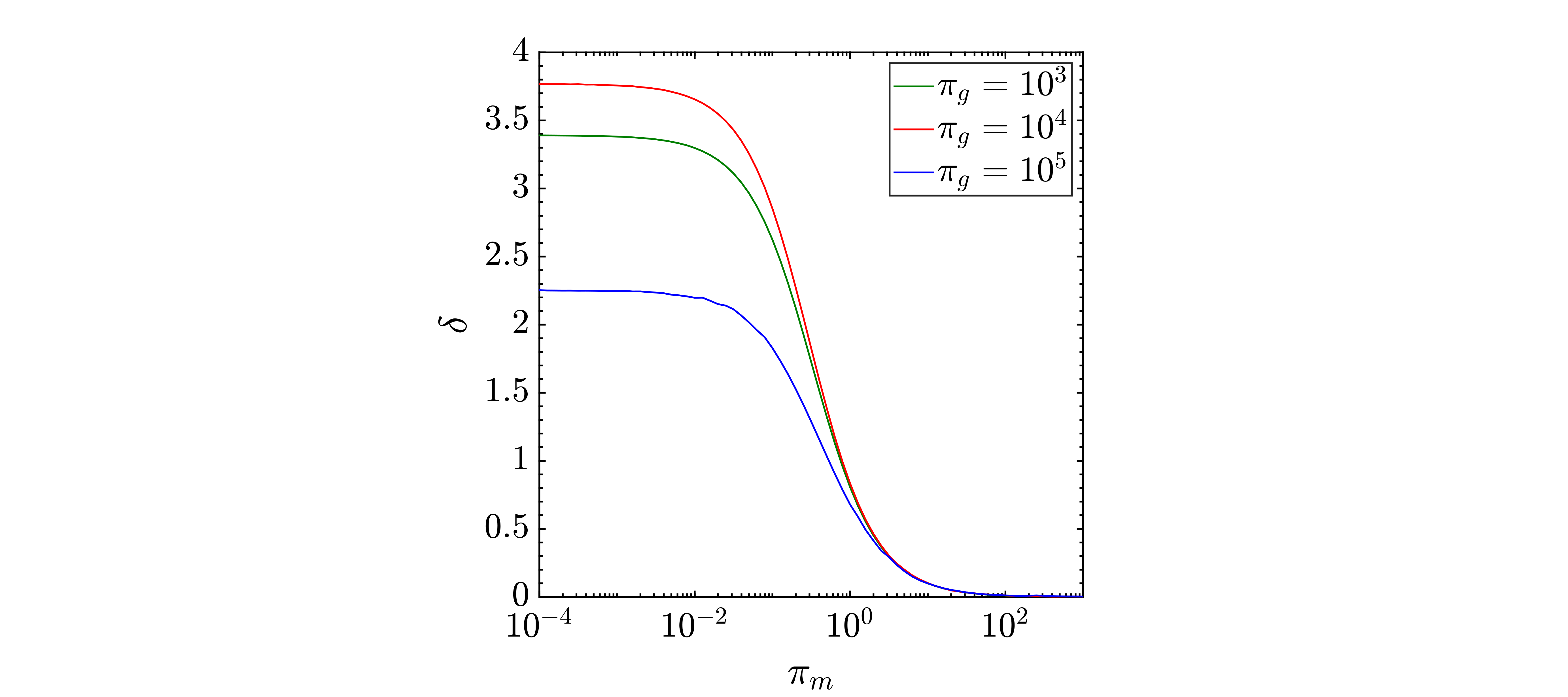}}
\subfloat[(d)]{\includegraphics[width=0.5\textwidth,trim={13.5cm 0cm 13.5cm 0cm},clip]{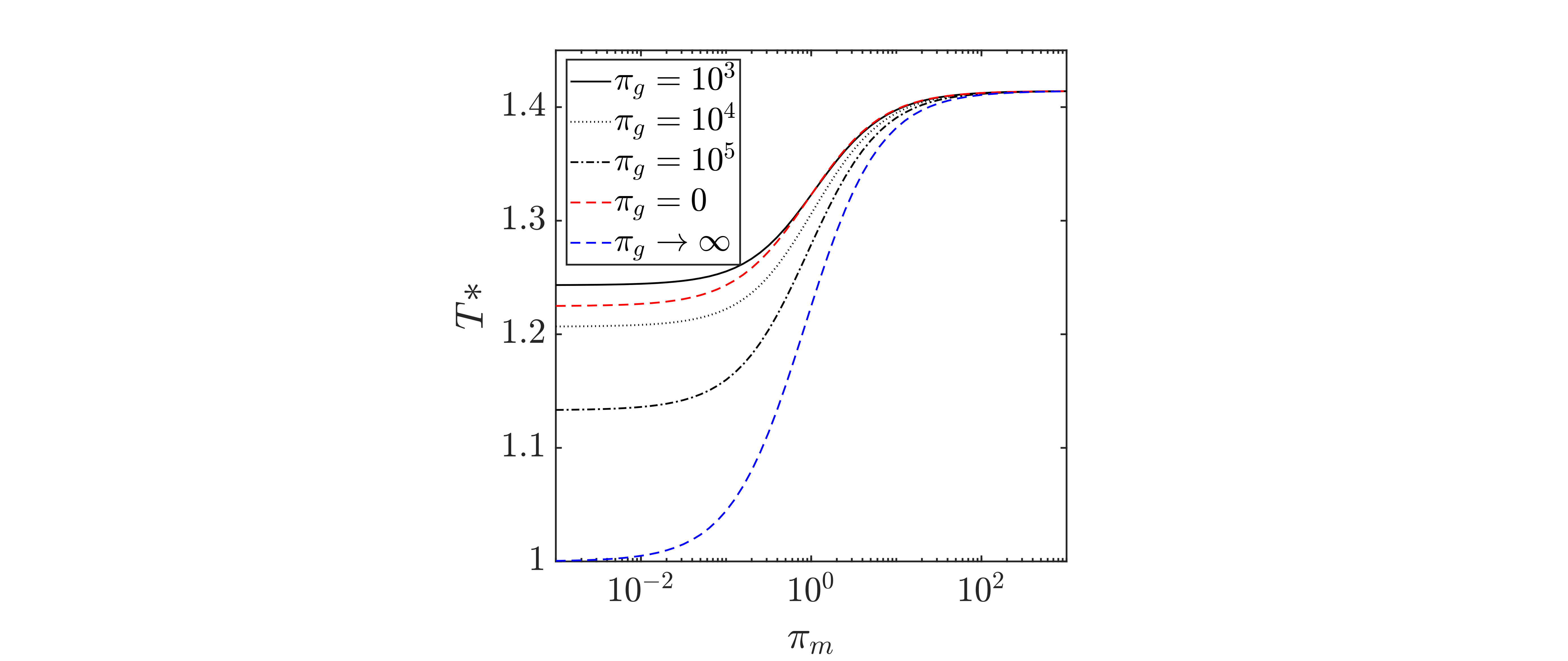}}

\caption{Figures (a) and (b) show the variation of $\delta$ and $T^*$ with $\pi_g = \frac{g R^3}{\nu^2}$ at different values of $\pi_m$. $\delta \to 0$ for both $\pi_g \to 0,\infty$ in (a). Also, $\delta$ is maximum for $\pi_g \approx 3980$ and is nearly invariant for all $\pi_m$. In (b), the locus of the point of maximum deviation is indicated with a red dotted curve. Figures (c) and (d) depict their variation versus $\pi_m = \frac{M}{m}$ for different values of $\pi_g$.} 
\label{pim-pig}
\end{center} 
\end{figure}




Figure \ref{pim-pig} is a plot of $\delta$ and $T^*$ versus $\pi_g$ and $\pi_m$. Figure \ref{pim-pig}(a) is a plot of $\delta$ versus $\pi_g$ for three different values of $\pi_m$. $\delta$ varies non-monotonically with $\pi_g$. For the case considered in this figure with a terminal point at $(10,-1)$ and $\pi_m=0.1$, the maximum deviation occurs at $\pi_g\approx 3981$. Figure \ref{pim-pig}(b) is a plot of $T^*$ versus $\pi_g$ for varying $\pi_m$. It can be seen from this plot that $T^*$ increases first and then decreases as $\pi_g$ increases.The reason for this behavior lies in the fact that at both low and high viscosity cases, the dissipation due to the fluid vanishes. This implies that a fluid bottle in both these limits will be faster than at other intermediate $\pi_g$ values. As $\pi_m$ increases, $T^*$ becomes insensitive to $\pi_g$.

Figure \ref{pim-pig}(c) is a plot of $\delta$ versus $\pi_m$ for three  values of $\pi_g$. As $\pi_m$ increases, the mass of the fluid is decreasing in relation to the bottle mass. As expected, $\delta$ decreases and tends to $0$ as $\pi_m \to \infty$. The variation of the geometric norm $(\delta)$ with respect to $\pi_m$ shows a sigmoid-like behavior with an inflection point at $\pi_m\approx 1$. In the low viscosity limit, the fluid bottle is effectively a hoop with mass $M$ and a point mass with mass $m$ at the center. Whereas, in the high viscosity limit, the fluid bottle acts like a hoop of mass $M$ around a solid cylinder of mass $m$. Therefore, in these two limits, dynamics of the fluid bottle will reduce to rigid body dynamics which leads to the brachistochrone being the cycloid itself. This is the reason for the geometric norm $\delta$ to vanish in these two limits (high and low viscosity). Figure \ref{pim-pig}(d) is a plot of $T^*$ versus $\pi_m$ for the same three values of $\pi_g$. Again, as $\pi_m$ increases, $T^*$ increases and reaches an asymptotic limit of $\sqrt{2}$. The asymptotic value of $T^*$ is square root of the ratio of the net inertial mass of a hoop to that of a particle, which is $\sqrt{2}$. See Appendix \ref{labelA} for a mathematical discussion of the origin of this limiting value.

\begin{figure}
\begin{center}
\centering
\captionsetup[subfigure]{labelformat=empty,labelsep=none}

\includegraphics[scale=0.2,trim={5cm 0cm 45cm 0cm},clip]{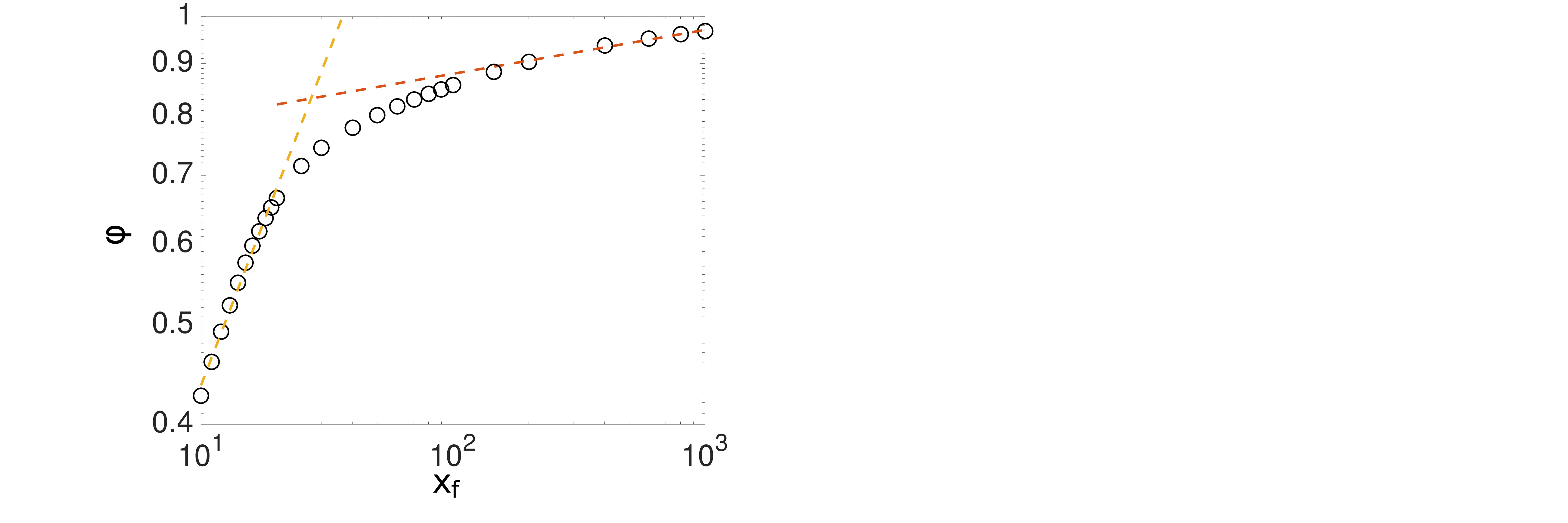}

\caption{Plot of dissipation fraction ($\phi$) versus $x_f$ for $y_f=-1$, $\pi_g=1000$ and $\pi_m=0.01$. Dotted lines indicate the two asymptotic regimes. For small $x_f$, $\phi \sim x_f^{0.64}$ and for large $x_f$, $\phi \sim x_f^{0.04}$.} 
\label{xf-diss}
\end{center} 
\end{figure}
\subsection{Role of viscous dissipation}
The total initial available energy (in dimensional terms) in the bottle-fluid system is $(M+m)gy'_f$ in all cases. As $x_f$ increases, the brachistochrone is observed to deviate from a cycloid (see Figure \ref{fig2}(b)). We will argue herein that this is due to increasing total dissipative loss and show the consequences thereof. The shape of the optimal curve is determined by a competition between two objectives - (i) the desire to minimize travel time ($t_f$) and (ii) the need to (at least) reach the end point in the face of viscous losses. This competition is best quantified in terms of the variation of the viscous dissipation as a function of end point coordinate $x_f$. We define the total dissipation $\Phi$ as
\begin{equation}
\Phi=\int_0^{t_f} \mu {\left[ r \frac{\partial}{\partial r}\left( \frac{v}{r}\right)\right]}^2 dt,
\end{equation}
\noindent which is the total energy dissipated during the travel time $t_f$. We then calculate $\phi=\frac{\Phi}{|(M+m)gy'_f|}$ which represents a fraction of the initial potential energy that has been dissipated by viscous action. Figure \ref{xf-diss} is a plot of $\phi$ as a function of $x_f$ with $y_f$ set equal to $-1$. Two distinct asymptotic regimes can be discovered for short and long $x_f$. For $x_f < 30$, $\phi \sim x_f^{0.64}$. This is the regime where the system initial potential energy is more than sufficient to ensure that the bottle reaches the end point. Therefore, the brachistochrone is primarily governed by the time-minimization objective. On the other hand, for $x_f > 100 $, $\phi \sim x_f^{0.04}$ implying that $\phi$ is a much weaker function of $x_f$. In this regime, the brachistochrone is also governed by a desire to conserve the initial potential energy in order to ensure that the bottle reaches the end point, while still minimizing travel time. Under this action, the time-minimization objective takes a back seat in favor of reaching the end point. This is the reason for the two different slopes at the two asymptotic regimes in Figure \ref{xf-diss}. To draw an analogy to running, the bottle chooses to `sprint' for short $x_f$, while for long $x_f$, the bottle is trying to run a `marathon'. The shapes of the brachistochrone curves under these conditions are presented in Figure \ref{fig2}. The transition between the two states - where the brachistochrone is nearly a cycloid and where the brachistochrone deviates from a cycloid - happens near $x_f \approx 30$ in this case (which would be a function of $y_f$ and other system parameters). In conclusion, the fluid brachistochrone problem is significantly more interesting owing to the fact that viscous dissipation plays a non-monotonic and important role. 

Several interesting extensions of this study can be explored. For instance, \cite{aristoff2009elastochrone} introduced elastic forces of the inclined surface and discussed the motion of a rigid cylinder on such a straight deformable incline. Also, \cite{balmforth2007dissipative} discussed the dissipative descent of a compound object in a gravitational field, again on a straight inclined plane. It would be interesting to explore brachistochrone versions of these problems.

\section{Conclusion}

We calculate the brachistochrone for the motion of a fluid-filled cylinder in a uniform gravitational field as a function of the properties of the fluid as well as the terminal points $(x_f,y_f)$. We define $\pi_g$ and $\pi_m$ as two dimensionless parameters governing the shape of the brachistochrone. $\pi_g$ is a ratio of the gravitational to viscous forces and $\pi_m$ is a ratio of the mass of the bottle to that of the fluid. We show that for $\pi_g \to 0,\infty$ as well as for $\pi_m \to \infty$, a cycloid is a good approximation of the brachistochrone. Firstly, we find that the effects of fluid on the deviation from a cycloid are significant only when $\pi_m<1$ i.e., the fluid mass is greater than the mass of the bottle. We find that the maximum deviation from a cycloid occurs for $\pi_g\approx 10^3$ and $\pi_m \to 0$  for a wide range of terminal point conditions. For small $x_f$, the brachistochrone is nearly coincident with a cycloid. Under this condition, the system is not constrained by viscous dissipation to achieve a time-minimized solution. As $x_f$ increases, the bottle chooses to operate in a mode which attempts to reduce viscous losses. Under this condition, the objective is dominated by a need to reach the end condition while still minimizing travel time. All of the above analyses assume laminar flow inside the cylinder and is rigorously valid only in the high viscosity (low Reynolds number) regime. For the low viscosity regime of $\pi_g \to \infty$, this analysis can at best be treated as a heuristic.

An interesting application of this work would be to study the brachistochrone motion of liquid drops on a super-hydrophobic surface as has been recently reported by \citet{tracks} and \cite{dropsbrac}. Interestingly, \cite{dropsbrac} compared the time of travel of a liquid droplet on a cycloid and a straight line. In addition, as discussed by \cite{janardan2015liquid}, liquid marbles are interesting fluid dynamic objects. Their brachistochrone motion in a gravitational field (ignoring drop deformation) is an example application where this work has practical implications. For example, a $2~mm$ diameter marble of a fluid whose kinematic viscosity is $10^{-5}~m^2s^{-1}$ corresponds to $\pi_m \approx 0$ and $\pi_g \approx 10^3$. For such a realistic drop, the deviation from cycloid is significant and the framework described here provides a way to compute the brachistochrone.



\bibliographystyle{jfm}
\bibliography{citations}

\appendix

\section{}\label{labelB}

In this appendix, we provide a brief derivation of the extension of Batchelor's solution for the case of a fluid flow inside a rotating cylinder originally due to \cite{supekar2014dynamics}. This appendix is a summary of their work being presented for completeness. We begin with the solution for the case where, an initially quiescent fluid is suddenly set to motion by the cylindrical shell which is imparted a constant angular velocity of value $\Omega_0 = 1/R$. The equations governing the fluid flow for this case are the same as \eqref{ge} with a boundary condition $v(R,t)= 1$ and an initial condition $v(r,0) = 0$. An analytical solution was given by \cite{batchelor} as follows
\begin{equation}
v(r,t) = 
  \frac{r}{R}+2 \sum_{n=1}^{\infty}\frac{J_1\left(\lambda_n\frac{r}{R}\right)}{\lambda_n J_0(\lambda_n)}
\exp{\left(-\lambda_n^2 \frac{\nu (t)}{R^2}\right)}.
\end{equation}
To extend this solution to the case where, the boundary condition is time varying, we invoke Duhamel's Theorem as stated by \cite{duhamel}. \newline
\begin{displayquote}
If $T_f(x,t)$ is the response of a linear system to a
single, constant non-homogenous term with
magnitude unity, then the response of the same
system to a single, time-varying non-homogenous
term with magnitude $B(t)$ can be obtained from the
following expression
\end{displayquote}

\begin{equation}
    T(x,t) = B(0)\ T_f(x,t) + \int_{\tau=0}^t T_f(x, t-\tau) \frac{d B(\tau)}{d t} d\tau.
\end{equation}
This set of equations \eqref{ge} is linear and hence, Duhamel's theorem is applicable to the system of a fluid-filled cylinder. Now, considering a boundary condition of the form $v(R,t) = \Omega(t) R$ and applying Duhamel's theorem, we get a closed form solution for the fluid flow as follows
\begin{equation}
v(r,t) = 
\int_{\tau=0}^{t}\left( r+2R\sum_{n=1}^{\infty}\frac{J_1\left(\lambda_n\frac{r}{R}\right)}{\lambda_n J_0(\lambda_n)}
\exp{\left(-\lambda_n^2 \frac{\nu (t-\tau)}{R^2}\right)}
\right) \dot{\Omega}(\tau) d\tau.
\label{b3}
\end{equation}
It is to be noted that the above solution is valid even in the case when the boundary condition evolves along with the flow. In case of a fluid-filled cylinder rolling down an arbitrary curved incline as shown in figure \ref{fig:schematic}, the dynamics governing the boundary condition i.e., the angular velocity of the cylindrical shell is given by
\begin{subequations}
\begin{align}
(m+2M) R^2\dot{\Omega}(t) = -T(t) + (m+M) g R sin(\gamma(t)).
\intertext{Using \eqref{b3}, we compute the torque on the cylindrical shell due to the viscous drag as}
T(t) = 4\pi \nu \rho R^2 h \sum_{n=1}^{\infty}{\int_{\tau = 0}^{t}{exp{\left(-\lambda^2_n\frac{\nu (t-\tau)}{R^2}\right)}\dot{\Omega}(\tau)}}d\tau.
\end{align}
\label{evolve}
\end{subequations}
\noindent Here, $\gamma(t)$ is the instantaneous angle of inclination encountered by the cylinder, $m$ is the mass of the fluid, $M$ is the mass of the cylindrical shell and $g$ is the acceleration due to gravity. It can be observed that if $\gamma(t)$ is known, equations \eqref{b3} and \eqref{evolve} form a closed system and hence can be solved numerically if not analytically. Hence, $\gamma(t)$ would be a convenient control parameter that could to be varied in order to optimize the inclined curve for minimizing the travel time of the fluid-filled cylinder.

\section{}\label{labelA}
In this appendix, we'll provide a brief mathematical derivation of the brachistochrone for the asymptotic cases of the fluid-filled cylinder. Since the scales for non-dimensionalization won't hold at these asymptotes, we go back to the dimensional form of the governing equations given in equations \eqref{VelocityEquation} and \eqref{TorqueEquation}.\\
\newline
\noindent \textbf{Case 1:}
\newline
Consider the case where $\nu \to 0$. In non-dimensional terms, $\pi_g\to\infty$. In this case, equations \eqref{TorqueEquation} are reduced to
\begin{equation}
\lim_{\nu \to 0} T(t) = 4\pi \rho R^2 h \sum_{n=1}^{\infty}{\int_{\tau = 0}^{t}\lim_{\nu \to 0} \nu\ {exp{\left(-\lambda^2_n\frac{\nu (t-\tau)}{R^2}\right)}\dot{\Omega}(\tau)}}d\tau.
\end{equation}
\noindent Hence,
\begin{equation}
\lim_{\nu \to 0} T(t) = 0.
\end{equation}
At this limit, the fluid inside doesn't rotate but slips against the wall of the cylinder. This is shown by the fact that the rotational velocity of the fluid inside the cylinder vanishes at every radial location:
\begin{equation}
\lim_{\nu \to 0} v(r,t) = 
\int_{\tau=0}^{t}\left( r+2R\sum_{n=1}^{\infty}\frac{J_1\left(\lambda_n\frac{r}{R}\right)}{\lambda_n J_0(\lambda_n)}
\lim_{\nu \to 0} \exp{\left(-\lambda_n^2 \frac{\nu (t-\tau)}{R^2}\right)}
\right) \dot{\Omega}(\tau) d\tau.
\end{equation} 
\begin{equation}
\implies \lim_{\nu \to 0} v(r,t) = \int_{\tau=0}^{t}\left( r+2R\sum_{n=1}^{\infty}\frac{J_1\left(\lambda_n\frac{r}{R}\right)}{\lambda_n J_0(\lambda_n)}\right) \dot{\Omega}(\tau) d\tau.
\end{equation}
On computing the integrand explicitly using Mathematica, we find that the integrand converges to zero. Hence, the velocity field vanishes at each radial location at all times.
\begin{equation}
\lim_{\nu \to 0} v(r,t) = 0.
\end{equation}
\noindent For the corresponding non-dimensional regime of $\pi_g \to \infty$, the torque applied by the fluid on the bottle vanishes. Since the fluid is slipping against the rotating cylinder, the fluid is equivalent to a point mass at the center of the cylinder. Hence, at this limit, the fluid-filled cylinder is effectively a hoop of mass $M=\pi_m$ with a point mass of mass $m=1$ at the center.\\

\noindent \paragraph{\textbf{Case 2:}}\label{2}
\newline
Consider the case where $\nu \to \infty$. In non-dimensional terms, it means $\pi_g\to 0$. In this case, equations \eqref{TorqueEquation} become
\begin{equation}
\lim_{\nu \to \infty} T(t) = 4\pi \rho R^2 h \sum_{n=1}^{\infty}{\int_{\tau = 0}^{t}\lim_{\nu \to \infty} \nu\ {exp{\left(-\lambda^2_n\frac{\nu (t-\tau)}{R^2}\right)}\dot{\Omega}(\tau)}}d\tau.
\end{equation}
Since exponential decay is faster than linear growth as $\nu \to \infty$,
\begin{equation}
\lim_{\nu \to \infty} T(t) = 0.
\end{equation}

\noindent Therefore, for the corresponding non-dimensional regime of $\pi_g \to \infty$, the torque applied by the fluid on the bottle vanishes. Unlike for the previous case, it isn't because of slipping of fluid against the cylinder. Here, it is because the fluid takes infinitesimally low transient time to reach the saturated velocity profile as shown below:

\begin{equation}
\lim_{\nu \to \infty} v(r,t) = 
\int_{\tau=0}^{t}\left( r+2R\sum_{n=1}^{\infty}\frac{J_1\left(\lambda_n\frac{r}{R}\right)}{\lambda_n J_0(\lambda_n)}
\lim_{\nu \to \infty} \exp{\left(-\lambda_n^2 \frac{\nu (t-\tau)}{R^2}\right)}
\right) \dot{\Omega}(\tau) d\tau.
\end{equation}
Again since exponential decay is faster than linear growth as $\nu \to \infty$, 
\begin{equation}
\lim_{\nu \to \infty} v(r,t) = r \Omega(t)
\end{equation}

\noindent Therefore, the fluid is undergoing a rigid body rotation. Hence, at this limit, the fluid-filled cylinder is equivalent to a hoop of mass $M=\pi_m$ and a solid cylinder of mass $m=1$. \\
\newline
\noindent \textbf{Brachistochrone for a general rigid body:}
\newline
Consider a rolling rigid body with a radius of gyration of $k$ and mass $M_b$. The brachistochrone for this rigid body is governed by the following variational problem
\begin{equation}
\minimize_{y(x) \in \mathbb{C(R)}} \ \sqrt{1+\frac{k^2}{R^2}}\int_{0}^{x_f} \sqrt{\frac{1+y'^2}{2 g y}} \ dx.
\end{equation}

\noindent Since the functional is a scalar multiple of that for a particle, the solution for the above variational problem is also a cycloid. Hence, for any rolling rigid body, for a minimum time travel under gravity, the center of mass must follow a cycloid.\\
\newline
Both the cases discussed earlier result in finding the brachistochrone of different rigid bodies. So, these cases lead to a brachistochrone which is a cycloid. Hence, at these asymptotes, the deviation from a cycloid vanishes. Also, for the general rigid body discussed above, the ratio of acceleration of the rigid body to that of a particle while on a cycloid curve is given by $1/(1+\frac{k^2}{R^2})$. Hence the ratio of time of descent i.e., the kinematic deviation($T^*$) as defined in equation \eqref{kdev} becomes $\sqrt{1+\frac{k^2}{R^2}}$. The radius of gyration for cases 1 and 2 discussed earlier are $k_1=\sqrt{\frac{\pi_m}{1+\pi_m}}R$ and $k_2=\sqrt{\frac{1+2\pi_m}{2(1+\pi_m)}}R$. This explains the asymptotic limits in the plot of $T^*$ vs $\pi_m$ as shown in figure \ref{pim-pig}(d) . Also, for the case of $\pi_m \to \infty$, the fluid-filled cylinder is effectively a hoop of mass $\pi_m$ with negligible mass of fluid. Hence, the radius of gyration of the effective rigid body is $k=R$. This means that the kinematic deviation tends towards $\sqrt{2}$ for $\pi_m \to \infty$ which is verified in figure \ref{pim-pig}(b).




\end{document}